\documentclass[12pt]{article}

\usepackage{graphicx}
\usepackage{epsfig}
\usepackage{amsmath}
\usepackage{mathrsfs}
\usepackage{amssymb}
\usepackage{amsthm}
\usepackage[authoryear]{natbib}
\usepackage{algorithm,algorithmic}
\usepackage{url}
\usepackage[margin=1in,paperwidth=8.5in,paperheight=11in]{geometry}
\usepackage{rotating}
\usepackage{setspace}
\usepackage{stmaryrd}
\usepackage{stackrel}
\usepackage{hyperref}
\usepackage{array,longtable}

\makeatletter
\newcommand\code{\bgroup\@makeother\_\@makeother\~\@makeother\$\@makeother\^\@codex}
\def\@codex#1{{\normalfont\ttfamily\hyphenchar\font=-1 #1}\egroup}

\newcommand\proglang{\bgroup\@makeother\_\@makeother\~\@makeother\$\@makeother\^\@codex}
\def\@codex#1{{\normalfont\ttfamily\hyphenchar\font=-1 #1}\egroup}

\newcommand\pkg{\bgroup\@makeother\_\@makeother\~\@makeother\$\@makeother\^\@codex}
\def\@codex#1{{\normalfont\ttfamily\hyphenchar\font=-1 #1}\egroup}
\makeatother

\newcommand{\sd}{\text{s.d.}}
\newcommand{\real}{\mathbb{R}}

\newcommand{\mcor}{\text{mcor}}
\newcommand{\eigen}{\mathrm{eigenvalues}}
\newcommand{\ecor}{\mathrm{cor}}

\begin{document}

\title{A Multi-Way Correlation Coefficient}
\author{Benjamin M. Taylor (Lancaster University)}
\maketitle

\begin{abstract}
     Pearson's correlation is an important summary measure of the amount of dependence between two variables. It is natural to want to generalise the concept of correlation as a single number that measures the inter-relatedness of three or more variables e.g. how `correlated' are a collection of variables in which non are specifically to be treated as an `outcome'? In this short article, we introduce such a measure, and show that it reduces to the modulus of Pearson's $r$ in the two dimensional case.
\end{abstract}

\section{Introduction and Literature Search}

Pearson's correlation coefficient, $r$, is applied in all areas of quantitative scientific endeavour, see \cite{stigler1989} for a review. For two samples $\{x_i\}_{i=1}^n$ and $\{y_i\}_{i=1}^n$, it is defined as:
\[r = \frac{\sum_{i=1}^n(x_i-\bar x)(y_i - \bar y)}{\sqrt{(x_i-\bar x)^2}\sqrt{(y_i-\bar y)^2}}.\]
In some instances, it is of interest to understand how related three or more variables are and there are a number of straightforward options in this case. Probably the most commonly applied of these would be:
\begin{enumerate}
    \item to examine the entries of the sample correlation matrix which summarise pairwise dependence, or
    \item to examine the coefficient of determination (a.k.a. the coefficient of multiple correlation), $R^2$, from a linear model assuming one variable as the outcome \citep{wright1921}.
\end{enumerate}
While both of these methods provide very useful summaries of the data, with the former method, the collection of plots can be difficult to interpret for even a moderate number of variables and the latter yields a different $R^2$ depending on the chosen outcome variable: which of these values should we choose to represent the correlation in our data? A secondary philosophical consideration with $R^2$ is that in some sense, one of the variables is being treated as an `outcome', which may in practice not be the case: there may be no justification for treating one of the variables in this way.

Why would this be useful? The motivation for this research occurred during a conversation with a Psychologist in which I was asked whether there was a way to measure dependence between three different variables capturing the properties of a certain type of brain signal: none of which was to be considered as an outcome. As another example, in a recent analysis of Tuberculosis incidence in two distinct areas in Portugal (in progress), we had occasion to examine the correlation between the covariates in our model to decide on whether there there was collinearity:

\begin{equation*}
    \left(\begin{array}{cccccc}
        1 & -0.13 & 0.18 & -0.27 & 0.19 & 0.36 \\
        -0.13 & 1 & 0.15 & 0.38 & 1\times10^{-1} & 0.11 \\
        0.18 & 0.15 & 1 & 0.15 & 8\times10^{-2} & 4\times10^{-2} \\
        -0.27 & 0.38 & 0.15 & 1 & -0.16 & -4\times10^{-2} \\
        0.19 & 1\times10^{-1} & 8\times10^{-2} & -0.16 & 1 & 0.14 \\
        0.36 & 0.11 & 4\times10^{-2} & -4\times10^{-2} & 0.14 & 1 \\
    \end{array}\right)
\end{equation*}

\begin{equation*}
    \left(\begin{array}{cccccc}
        1 & 0.23 & 0.51 & -7\times10^{-2} & 3\times10^{-1} & 0.58 \\
        0.23 & 1 & 0.12 & -3\times10^{-1} & 0.14 & 0.28 \\
        0.51 & 0.12 & 1 & -0.11 & 0.24 & 0.31 \\
        -7\times10^{-2} & -3\times10^{-1} & -0.11 & 1 & 2\times10^{-2} & 3\times10^{-2} \\
        3\times10^{-1} & 0.14 & 0.24 & 2\times10^{-2} & 1 & 0.33 \\
        0.58 & 0.28 & 0.31 & 3\times10^{-2} & 0.33 & 1 \\
    \end{array}\right)
\end{equation*}
As well as the specific 2-way correlations between variables in each of these areas, it is also of interest to ask the question: `are the measured covariates in one area more correlated than the other?' Our proposed measure gives an answer to that question.



There are many correlation coefficients in the literature. We used Web of Science to search titles for the words `correlation' and `coefficient' and restricted to the categories statistics probability or chemistry physical or engineering electrical electronic or engineering chemical or mathematics interdisciplinary applications; we did not place any restriction on the type of articles to search. This process yielded 1585 articles. We then searched each of these titles for the phrase `correlation coefficient', which left 700 articles. These were then manually categorised, either using the name of the correlation coefficient considered if it was explicitly stated in the title of the article, or classifying as `unnamed'; the results are in Table \ref{tab:corr_coef}.

None of these identified correlation coefficients is designed for the specific task we have in mind i.e. to describe, using a single number, the correlation between multiple realisations of a single $d$-dimensional random variable $d\geq2$.

\section{The Multi-Way Correlation Coefficient}

In order to circumvent the issues detailed above, we here propose a simple one-dimensional summary of linear inter-dependence in a $d$-dimensional real-valued random variable. We call this the \textbf{multi-way correlation coefficient} and define it as:
\[\mcor\left[\left(v_1,\cdots, v_n\right)\right] = \frac1{\sqrt{d}}\sd\left\{\eigen\left[\ecor\left(v_1,\cdots, v_n\right)^T\right]\right\},\]
where $v_i$ are column vectors containing the $d$-dimensional variables of interest for individual $i$ ($i=1,\ldots,n$), $\sd$ is the standard deviation and $\ecor$ is the empirical correlation matrix.

It is straightforward to see that when $d=2$, the multi-way correlation coefficient reduces to the modulus of Pearson's $r$. To see this, recall that $r$ features on the off-diagonal element of the correlation matrix, hence
\[\left|\begin{array}{cc}1-\lambda & r\\r & 1-\lambda\end{array}\right| = (1-\lambda)^2 - r^2.\]
Solving for zero yields eigenvalues $(1-r)$ and $(1+r)$; the empirical standard deviation of which is easily shown to be $\sqrt{2}r$. When $r$ is replaced by $-r$ in the correlation matrix, note the characteristic polynomial is still $(1-\lambda)^2 - r^2$, so the same argument follows for negative correlations, hence $\mcor\{x,y\} = |\ecor(x,y)|$ as claimed.

The multi-way correlation coefficient takes values on $[0,1]$. Let $\lambda_1,\ldots,\lambda_d$ be the eigenvalues of the correlation matrix. Then

\begin{eqnarray}
    0 \leq \mcor\left[\left(v_1,\cdots, v_n\right)^T\right] &=& \frac{1}{\sqrt{d}}\sqrt{\frac{1}{d-1}\sum_{i=1}^d(\lambda_i - \bar\lambda)^2} \\
    &=& \frac{1}{\sqrt{d}}\sqrt{\frac{1}{d-1}\left[\sum_{i=1}^d\lambda_i^2 -2\sum_{i=1}^d\lambda_i + d\right]}\\
    &=& \frac{1}{\sqrt{d}}\sqrt{\frac{1}{d-1}\left[\sum_{i=1}^d\lambda_i^2 - d\right]}\\
    &\leq& \frac{1}{\sqrt{d}}\sqrt{\frac{1}{d-1}\left[\left(\sum_{i=1}^d\lambda_i\right)^2 - d\right]} = 1
\end{eqnarray}
Where in the above, since the diagonal elements of a correlation matrix are all equal to $1$, we have used $\bar\lambda=\frac1d\sum_{i=1}^d\lambda_i=d/d=1$. The upper bound is attained when one eigenvalue is equal to $d$ and the rest equal to zero and the lower bound is attained when all eigenvalues are equal to 1 (corresponding to the identity matrix).

As is the case for for Pearson's correlation coefficient, if all variables are mutually independent, then the correlation matrix is the identity and the multi-way correlation coefficient will be equal to zero. A similar argument to that above shows that for an $n$-dimensional random variable, if $k$ of the components are independent from the rest then,
\[0 \leq \mcor\left[\left(v_1,\cdots, v_n\right)^T\right]\leq\left(\frac{(n-k)(n-k-1)}{n(n-1)}\right)^{1/2}.\]

The intuition behind the multi-way correlation stems from the eigendecomposition of the correlation matrix, $R$, say. Since $R$ is symmetric and positive definite, the eigenvectors form an orthonormal basis, which in some sense describe the axes of an ellipsoid that represents the multi-dimensional direction of dependence in the data. The eigenvalues scale these axes: a large eigenvalue (compared to the others) means the data is more `stretched' in the direction of the associated eigenvector, whereas if all eigenvalues are similar, then there is a similar amount of stretch in each direction. Hence a measure of spread of the eigenvalues gives a measure of how close to linear the data are. The particular choice of dispersion made in this article is due to the fact that in two dimensions it (nearly) reduces to Pearson's $r$; the concept of a `negative' correlation in $\real^d$, for $d\geq3$, is not meaningful, in any case.

As kindly pointed out by an anonmymous reviewer of an early version of the present article, there is similarity between our proposed measure of multi-way linearity and the measure of covariance sphericity of \cite{john1972}. For eigenvalues of a covariance matrix, $\lambda_1,\ldots,\lambda_d$, the covariance sphericity is defined as $\sum\lambda_i^2 / (\sum\lambda_i)^2$. Note that the denominator is equal to $d^2$ when this measure is applied to a correlation matrix. This measure takes values in $[d,1]$, so can be rescaled onto $[0,1]$ and used in a similar way to the measure we propose in the present article. Compared to this alternative, our proposed measure has the advantage of being in some sense equivalent to Pearson's $r$ in the two dimensional case.


\section{Examples}

Below are some examples of the multi-way correlation coefficient applied to 1000 random points in $\real^3$. Further examples are available in the web supplement.

\begin{figure}[H]
    \begin{minipage}{0.333\textwidth}

    \centering

    Two variables a linear function of the third.
    \begin{eqnarray*}
        x&\sim&\text{unif}(0,1)\\
        y&=&2x\\
        z&=&x
    \end{eqnarray*}

    \begin{tabular}{cccc}
          & x & y & z \\ \hline
        x & 1 & 1 & 1 \\
        y & 1 & 1 & 1 \\
        z & 1 & 1 & 1 \\
    \end{tabular}

    Eigenvalues: 3, 0, 0
    $\mcor\{x,y,z\}=1$
    \end{minipage}\begin{minipage}{0.333\textwidth}
        \includegraphics[width=\textwidth]{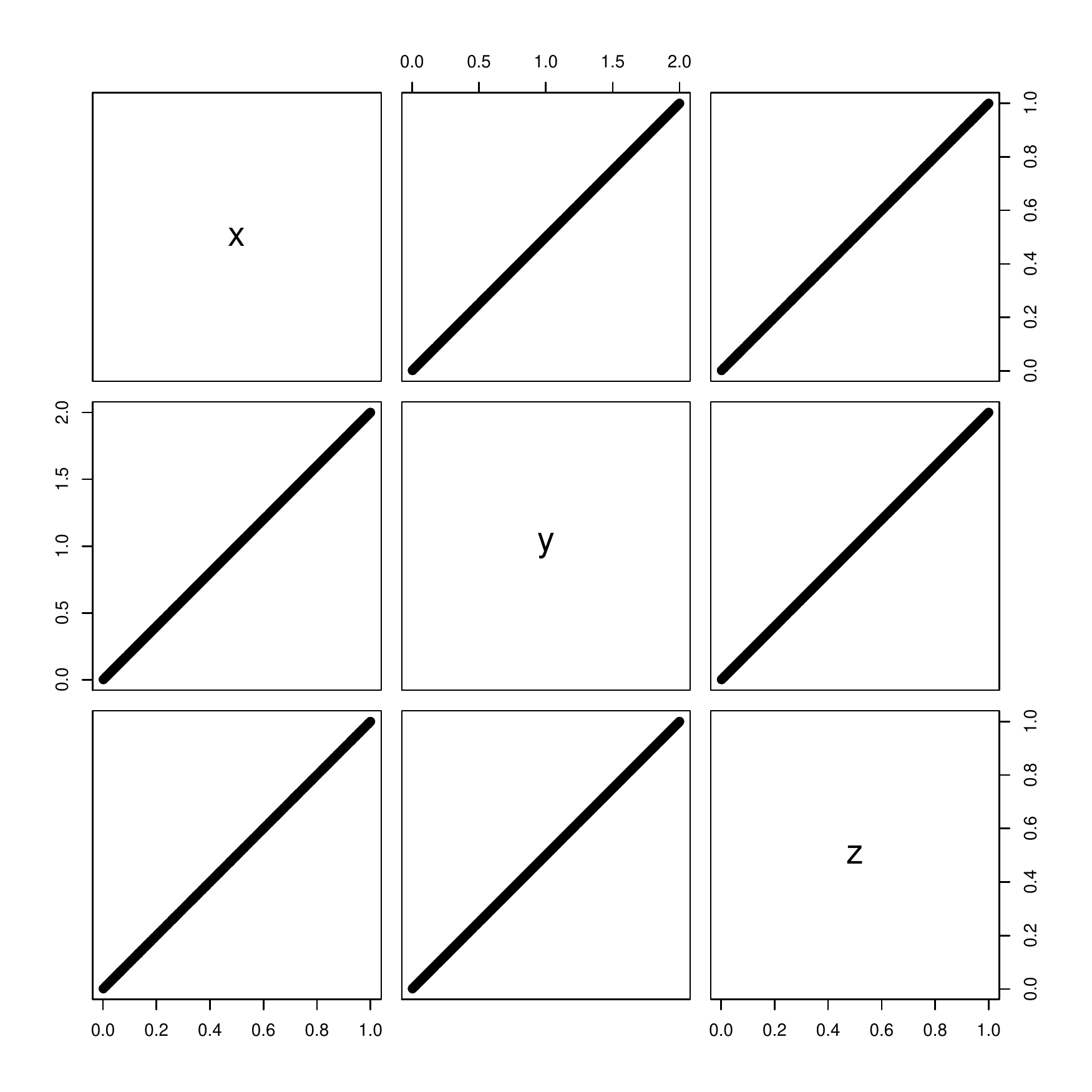}
    \end{minipage}\begin{minipage}{0.333\textwidth}
        \includegraphics[width=\textwidth]{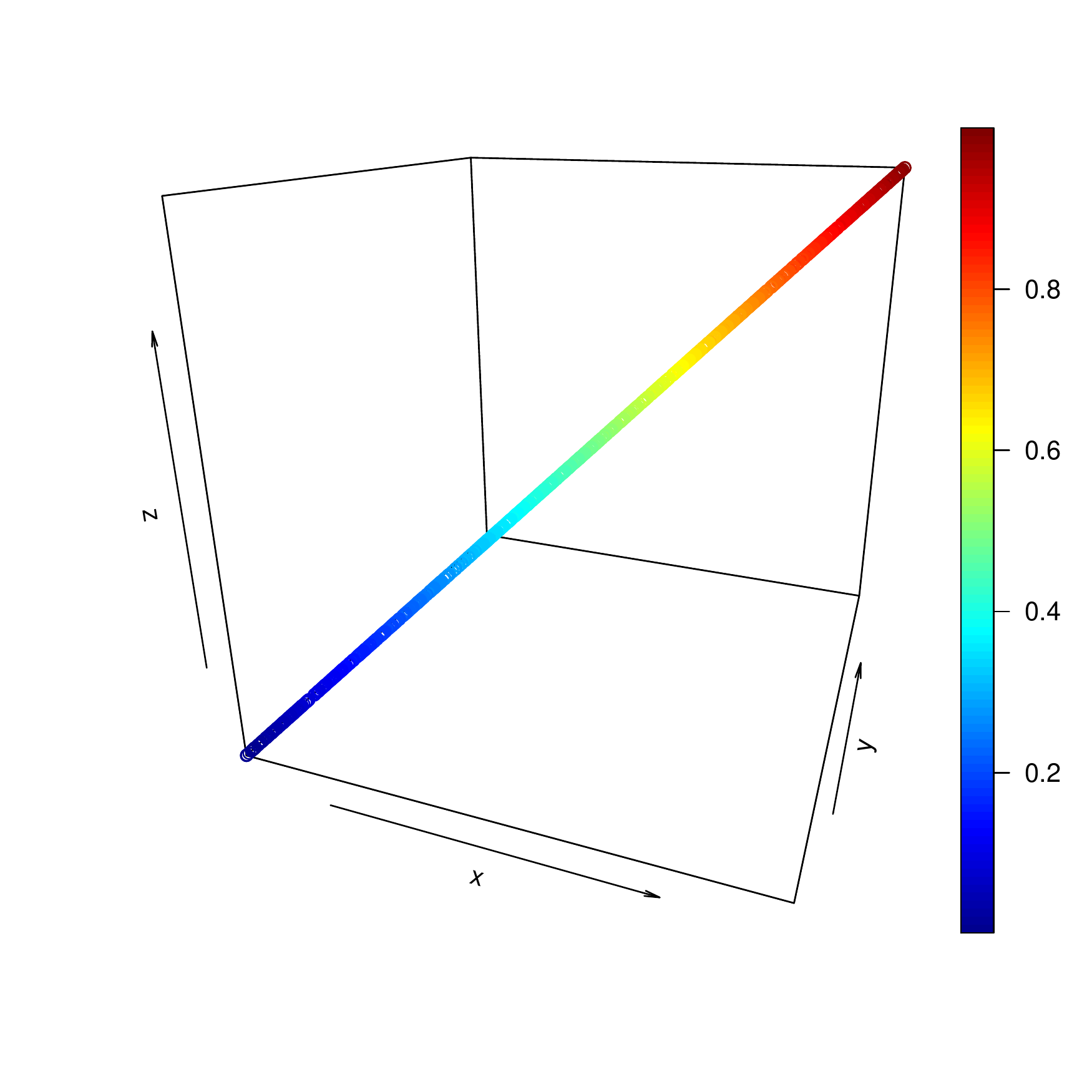}
    \end{minipage}
\end{figure}

\begin{figure}[H]
    \begin{minipage}{0.333\textwidth}

    \centering

    One variable a linear combination of other two.
    \begin{eqnarray*}
        x&\sim&\text{unif}(0,1)\\
        y&\sim&\text{unif}(0,1)\\
        z&=& x + 2y
    \end{eqnarray*}

    \begin{tabular}{cccc}
          & x & y & z \\ \hline
        x & 1 & -0.036 & 0.428 \\
        y & -0.036 & 1 & 0.888 \\
        z & 0.428 & 0.888 & 1 \\
    \end{tabular}

    Eigenvalues: 1.972, 1.028, 0
    $\mcor\{x,y,z\}=0.569$
    \end{minipage}\begin{minipage}{0.333\textwidth}
        \includegraphics[width=\textwidth]{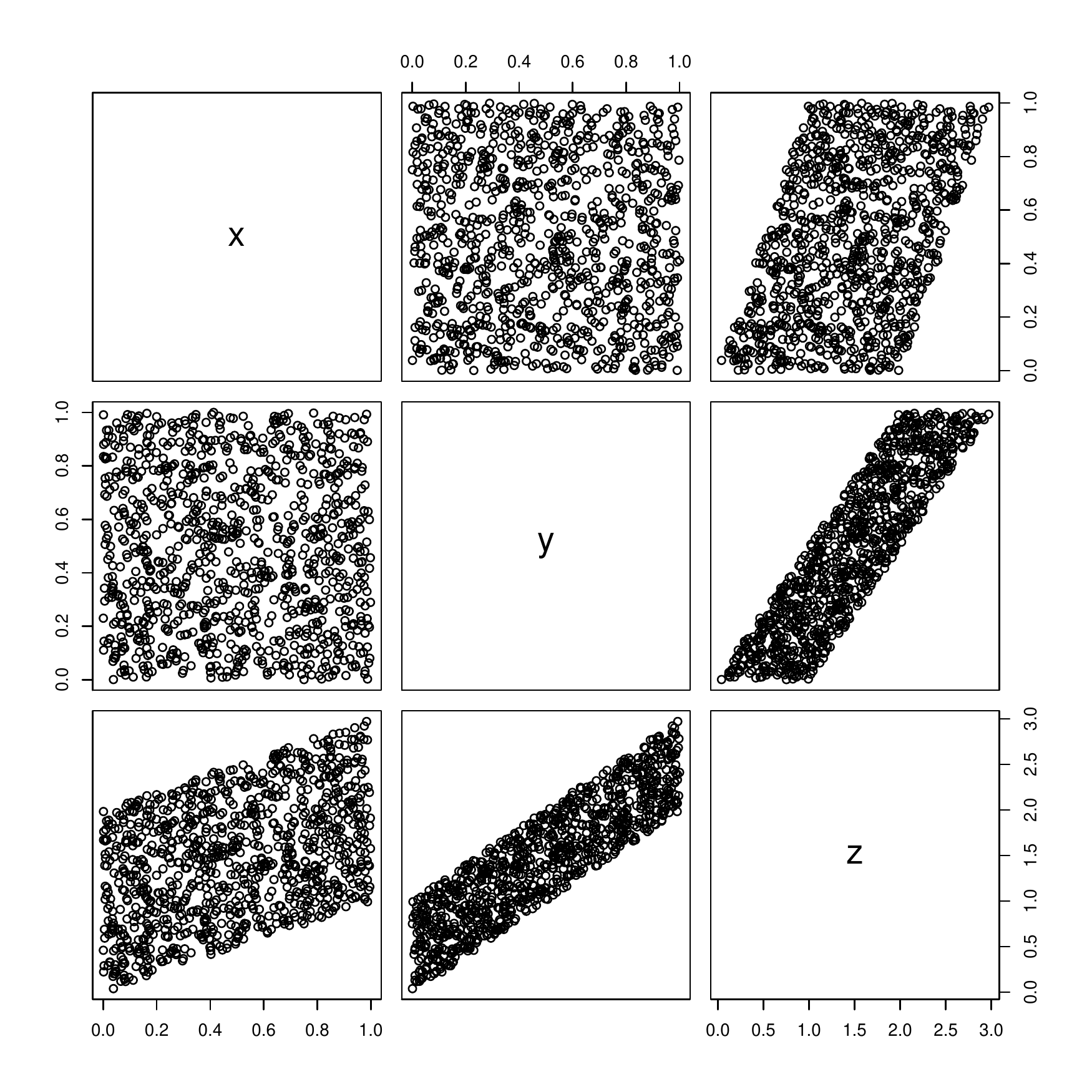}
    \end{minipage}\begin{minipage}{0.333\textwidth}
        \includegraphics[width=\textwidth]{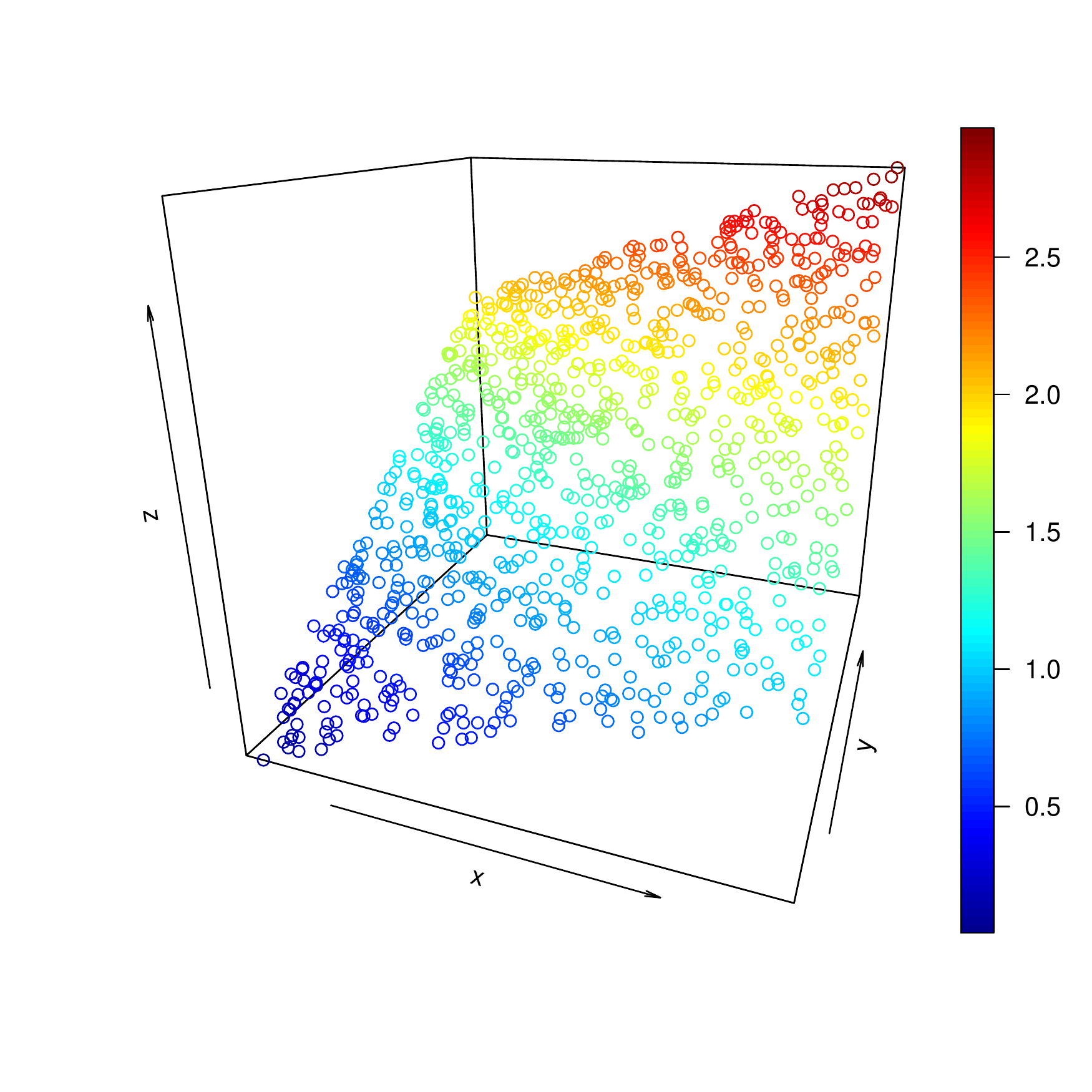}
    \end{minipage}
\end{figure}

\begin{figure}[H]
    \begin{minipage}{0.333\textwidth}

        \centering

    Three independent variables.
    \begin{eqnarray*}
        x&\sim&\text{unif}(0,1)\\
        y&\sim&\text{unif}(0,1)\\
        z&\sim&\text{unif}(0,1)
    \end{eqnarray*}

    \begin{tabular}{cccc}
          & x & y & z \\ \hline
        x & 1 & -0.026 & 0.024 \\
        y & -0.026 & 1 & 0.012 \\
        z & 0.024 & 0.012 & 1 \\
    \end{tabular}

    Eigenvalues: 1.03, 1.012, 0.958
    $\mcor\{x,y,z\}=0.021$
    \end{minipage}\begin{minipage}{0.333\textwidth}
        \includegraphics[width=\textwidth]{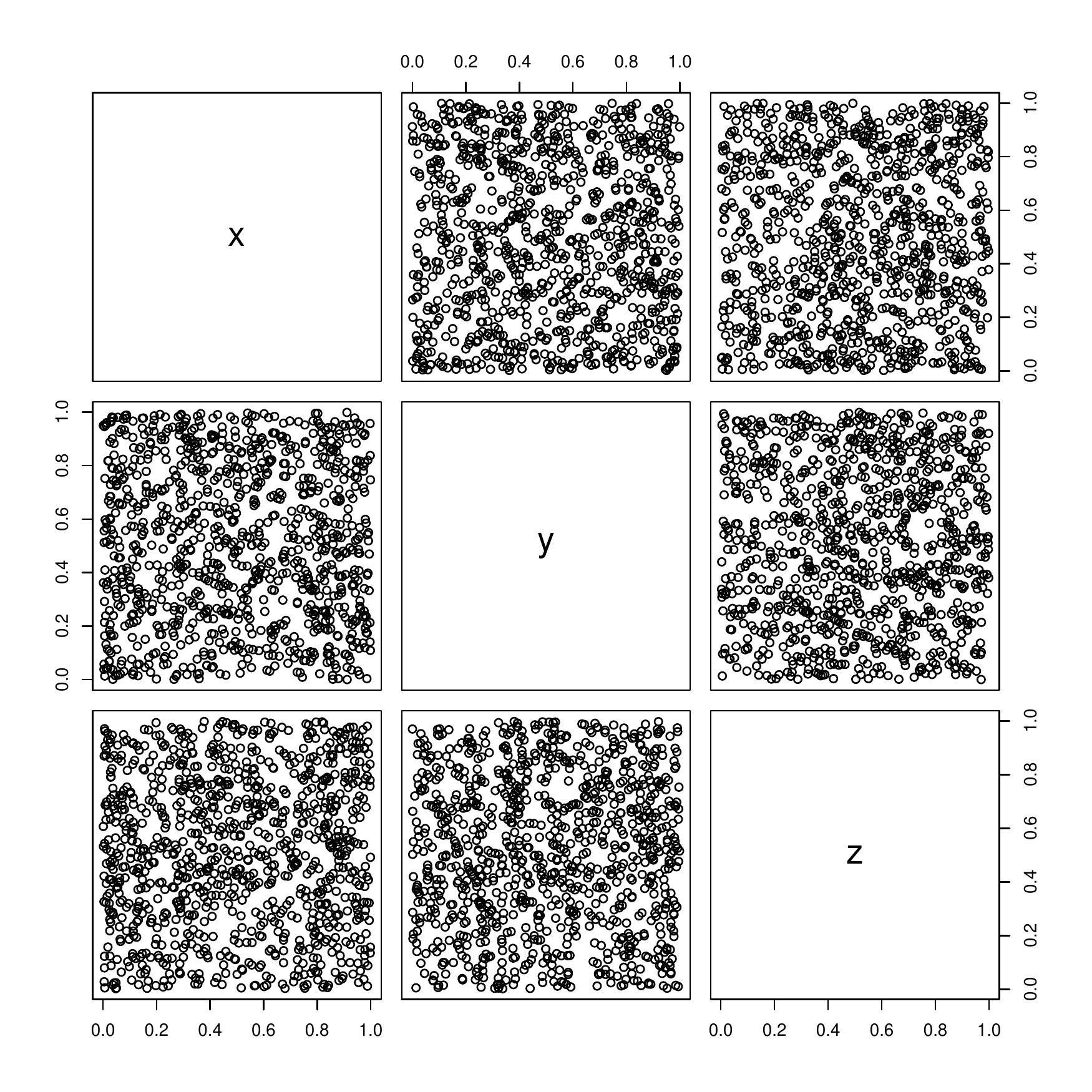}
    \end{minipage}\begin{minipage}{0.333\textwidth}
        \includegraphics[width=\textwidth]{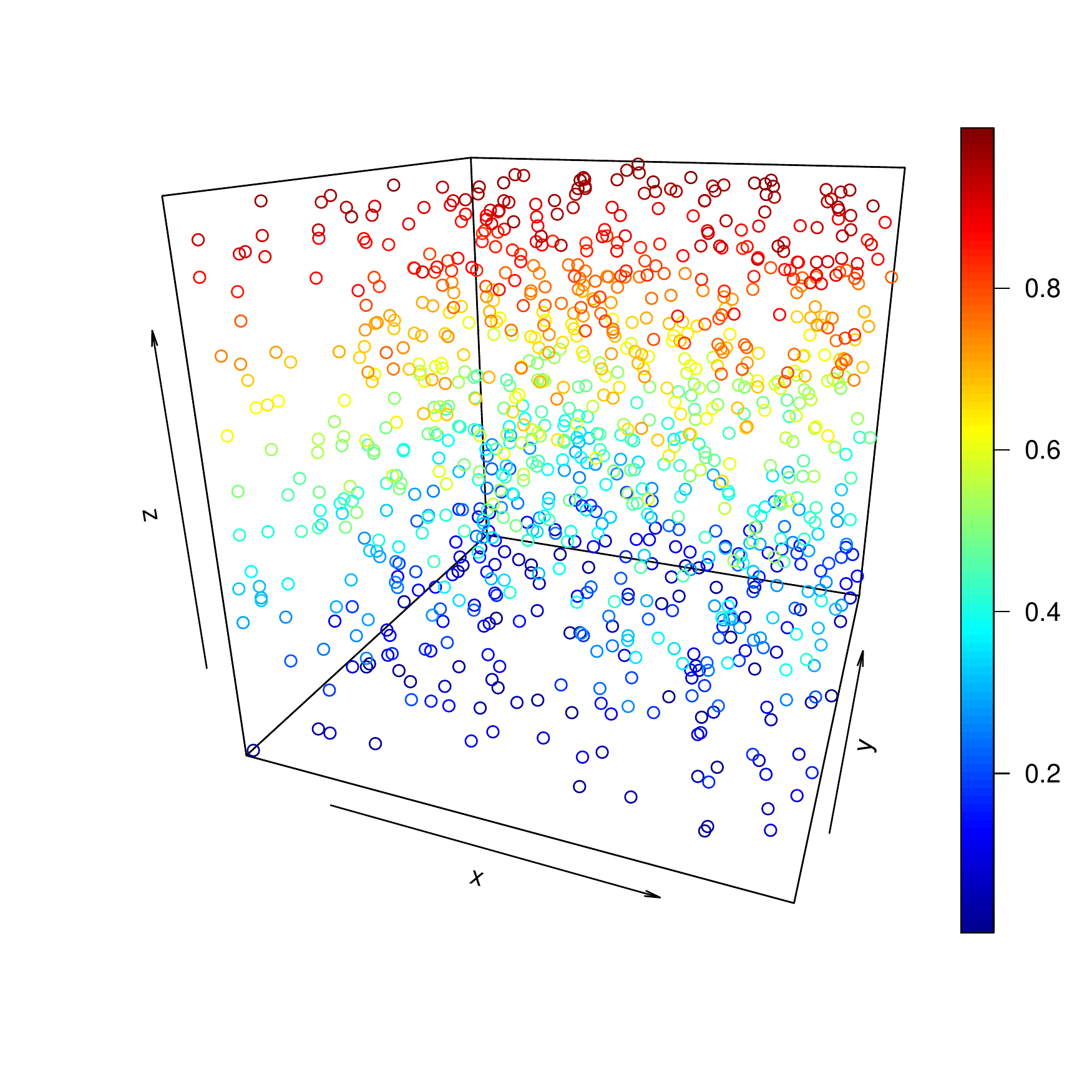}
    \end{minipage}
\end{figure}

\begin{figure}[H]
    \begin{minipage}{0.333\textwidth}

        \centering

    Correlated variables.
    \begin{eqnarray*}
        x&\sim&\text{unif}(0,1)\\
        y&\sim&\text{unif}(0,1)\\
        z&=& x + 2y + \text{N}(0,1)
    \end{eqnarray*}

    \begin{tabular}{cccc}
          & x & y & z \\ \hline
        x & 1 & -0.030 & 0.24 \\
        y & -0.030 & 1 & 0.46 \\
        z & 0.24 & 0.46 & 1 \\
    \end{tabular}

    Eigenvalues: 1.507, 1.024, 0.469
    $\mcor\{x,y,z\}=0.3$
    \end{minipage}\begin{minipage}{0.333\textwidth}
        \includegraphics[width=\textwidth]{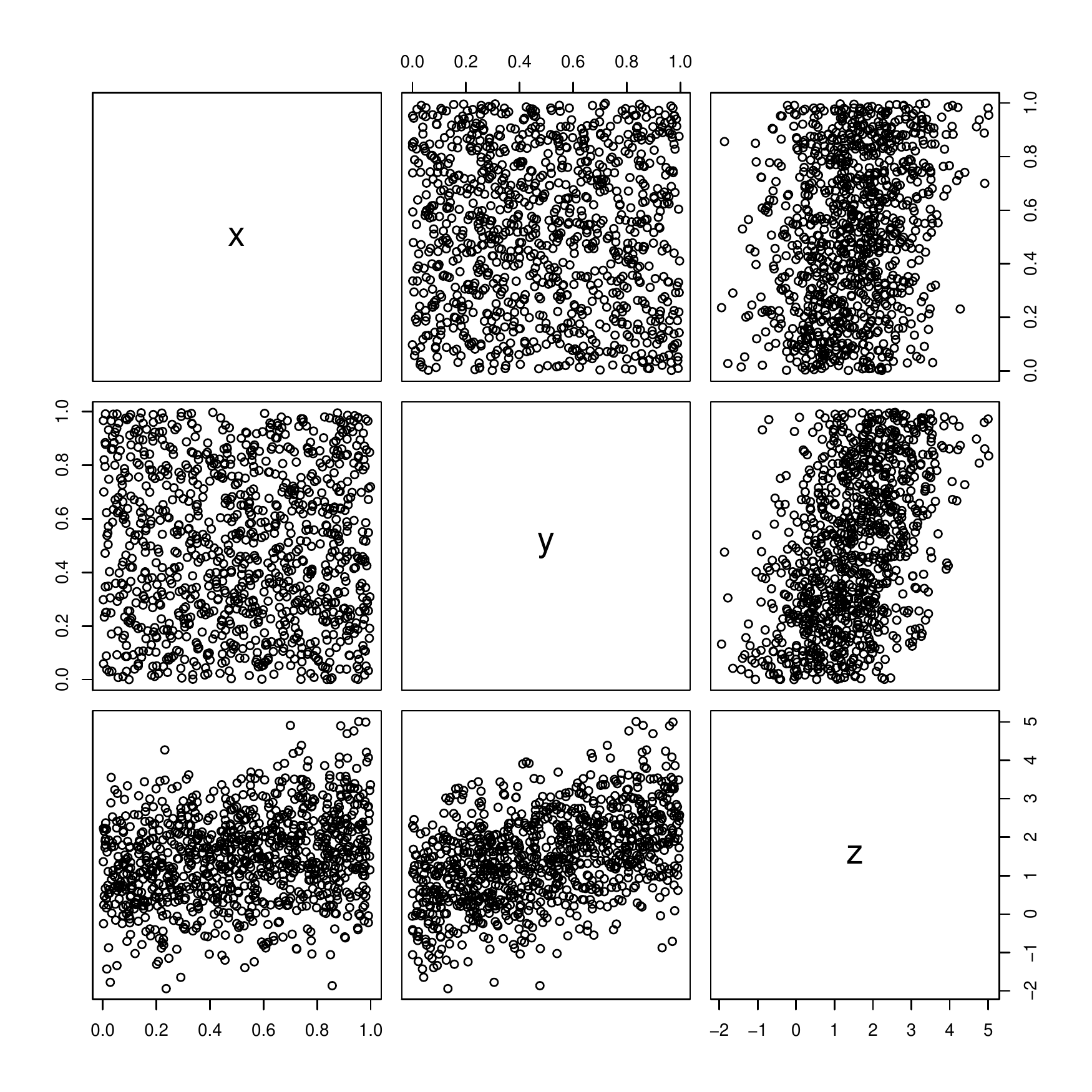}
    \end{minipage}\begin{minipage}{0.333\textwidth}
        \includegraphics[width=\textwidth]{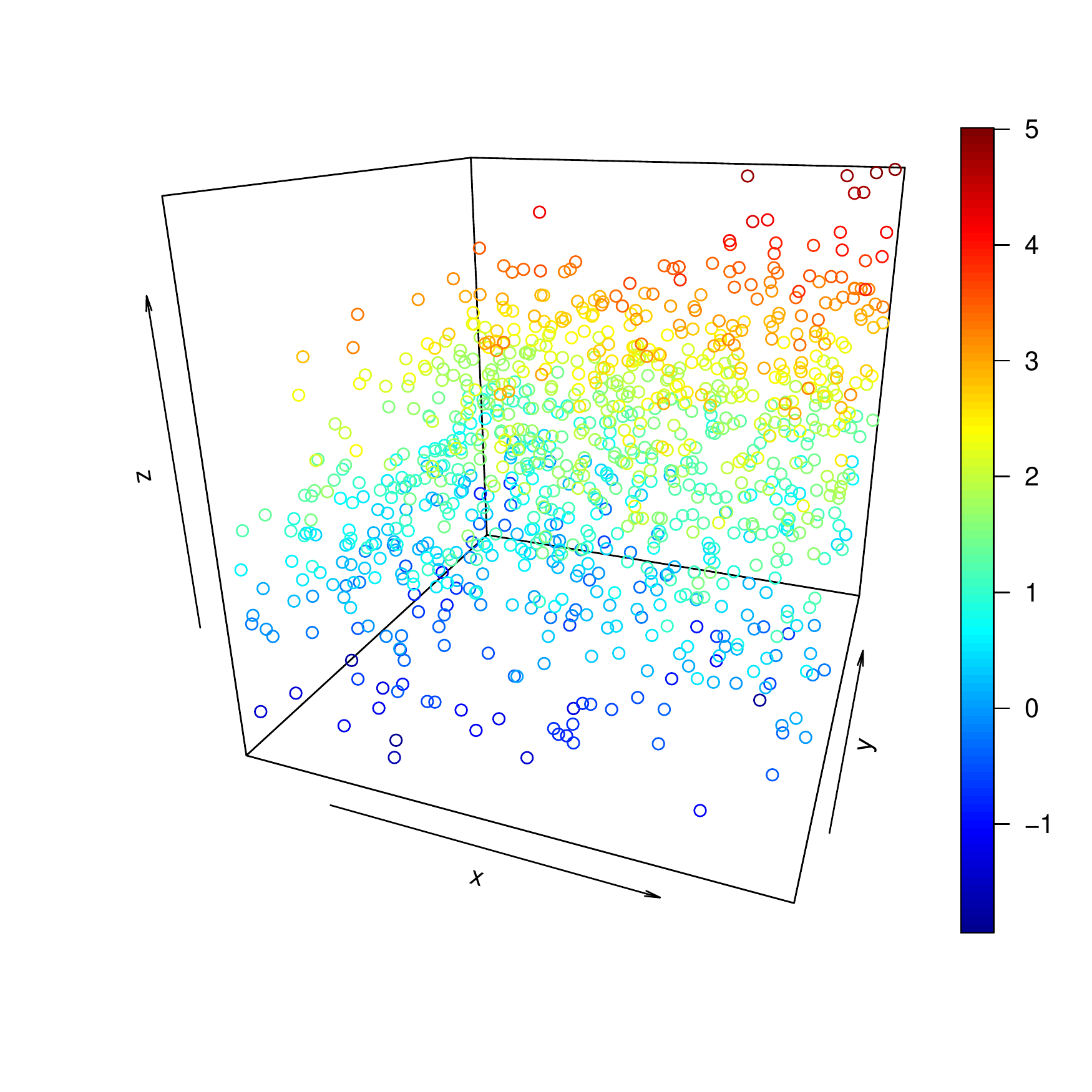}
    \end{minipage}
\end{figure}

\begin{figure}[H]
    \begin{minipage}{0.333\textwidth}

        \centering

    Even more Correlated variables.
    \begin{eqnarray*}
        x&\sim&\text{unif}(0,1)\\
        y&=& 5x + \text{N}(0,1)\\
        z&=& x + 2y + \text{N}(0,1)
    \end{eqnarray*}
    \begin{tabular}{cccc}
          & x & y & z \\ \hline
        x & 1 & 0.812 & 0.814 \\
        y & 0.812 & 1 & 0.966 \\
        z & 0.814 & 0.966 & 1 \\
    \end{tabular}

    Eigenvalues: 2.73, 0.236, 0.034
    $\mcor\{x,y,z\}=0.867$
    \end{minipage}\begin{minipage}{0.333\textwidth}
        \includegraphics[width=\textwidth]{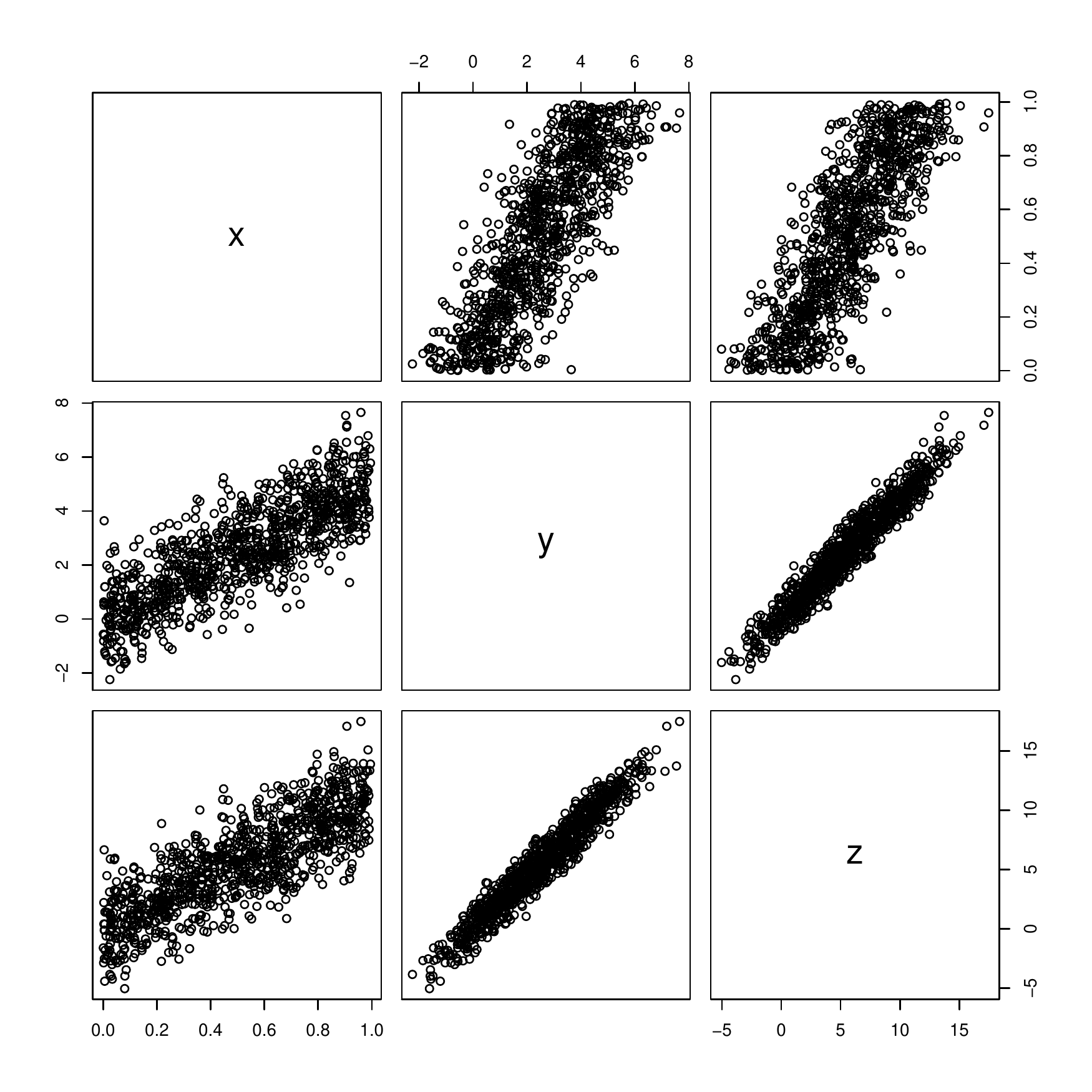}
    \end{minipage}\begin{minipage}{0.333\textwidth}
        \includegraphics[width=\textwidth]{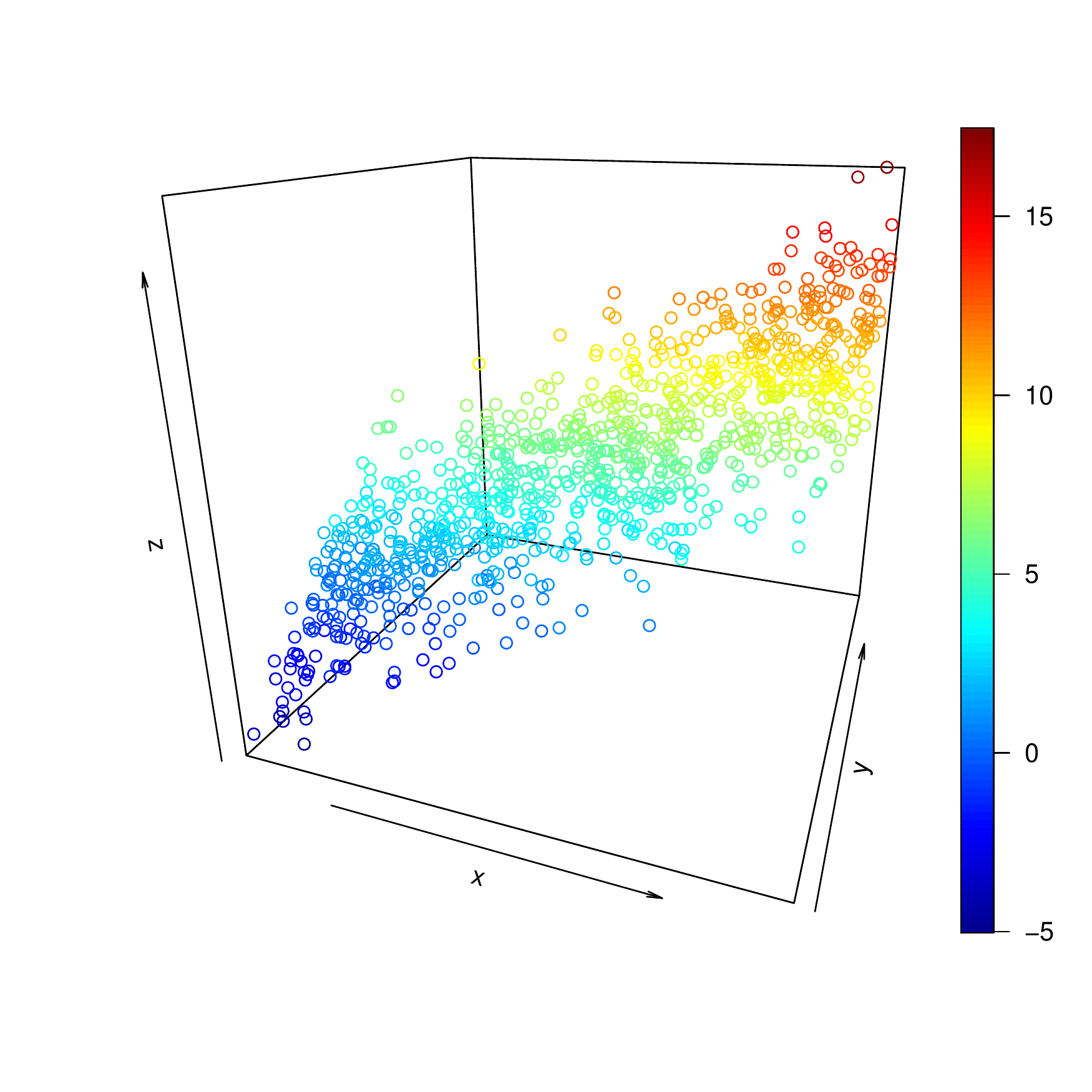}
    \end{minipage}
\end{figure}

\section{R Code}

R code for the multi-way correlation coefficient is straightforward:

\begin{verbatim}
    cor_taylor <- function(X){
        if(!inherits(X,"matrix")){
            stop("Input must be inherit 'matrix' class.")
        }
        n <- ncol(X)
        return((1/sqrt(n))*sd(eigen(cor(X))$values))
    }
\end{verbatim}

\section{Conclusion}

We have introduced a new correlation coefficient for $d$-dimensional random variables ($d\geq2$) that expresses in a similar sense to Pearson's correlation coefficient the amount of linear dependence between a set of variables. Our new measure is straightforward to compute and yields a 1 dimensional, easily interpretable, summary of dependence.

\bibliographystyle{chicago}
\bibliography{bibliography}

\appendix

\section{Table of Correlation Coefficients}

\begin{table}[htbp]
    \centering
    \tiny
    \begin{tabular}{lll|lll}
        Name & Range & $N$ & Name & Range & $N$ \\ \hline
        unnamed & 1946--2017 & 283 & regularised canonical & 2015--2015 & 1 \\
        intraclass & 1967--2016 & 54 & weighted rank & 2015--2015 & 1 \\
        multiple & 1952--2014 & 40 & zero-lag & 2015--2015 & 1 \\
        serial & 1945--2013 & 38 & polarization & 2014--2014 & 1 \\
        Pearson's & 1945--2016 & 33 & complex & 2013--2013 & 1 \\
        cross-correlation & 1962--2017 & 29 & information & 2013--2013 & 1 \\
        concordance & 1996--2015 & 23 & average & 2012--2012 & 1 \\
        tetrachoric & 1946--2010 & 15 & fuzzy & 2012--2012 & 1 \\
        spearman's & 1952--2012 & 12 & scale-invariant & 2012--2012 & 1 \\
        partial & 1946--2015 & 12 & sparse & 2011--2011 & 1 \\
        rank & 1947--2016 & 10 & bivariate normal & 2010--2010 & 1 \\
        canonical & 1969--2014 & 8 & multi-scale & 2010--2010 & 1 \\
        intra-cluster & 2001--2013 & 7 & possibilistic & 2010--2010 & 1 \\
        sample & 1966--2006 & 6 & rate-distortion & 2010--2010 & 1 \\
        envelope & 2007--2016 & 5 & weights-based & 2010--2010 & 1 \\
        generalised & 1996--2011 & 5 & covariate-adjusted & 2009--2009 & 1 \\
        normal & 2007--2016 & 4 & local & 2008--2008 & 1 \\
        antenna & 2012--2016 & 3 & non-parametric & 2008--2008 & 1 \\
        copolar & 2011--2017 & 3 & normalised & 2008--2008 & 1 \\
        order statistics & 2006--2008 & 3 & polarity coincidence & 2008--2008 & 1 \\
        semipartial & 2002--2010 & 3 & weighted & 2008--2008 & 1 \\
        auto-correlation & 1981--2013 & 3 & grade & 2007--2007 & 1 \\
        biserial & 1954--1954 & 3 & slant & 2007--2007 & 1 \\
        velocity & 2010--2010 & 2 & polychronic & 2006--2006 & 1 \\
        spatial & 2007--2013 & 2 & subshell-pair & 2006--2006 & 1 \\
        entropy & 2007--2011 & 2 & vector & 2006--2006 & 1 \\
        two-dimensional & 2006--2006 & 2 & circular polarization & 2004--2004 & 1 \\
        power & 2005--2010 & 2 & Gebelein's & 2004--2004 & 1 \\
        generalised concordance & 2001--2010 & 2 & robust & 2004--2004 & 1 \\
        polarimetric & 2001--2005 & 2 & skipped & 2004--2004 & 1 \\
        angular & 2001--2002 & 2 & regional & 2003--2003 & 1 \\
        e-correlation & 1978--1980 & 2 & amplitude & 2002--2002 & 1 \\
        uniform & 1971--1974 & 2 & frequency-correlation & 2002--2002 & 1 \\
        kendall's & 1970--2008 & 2 & sample serial & 2002--2002 & 1 \\
        partial auto-correlation & 1960--1983 & 2 & transformation & 2002--2002 & 1 \\
        circular serial & 1956--1960 & 2 & antiparallel electron & 1999--1999 & 1 \\
        genetic & 1955--1959 & 2 & dipolar & 1998--1998 & 1 \\
        zero-order & 1948--1948 & 2 & z-transformation & 1998--1998 & 1 \\
        distance & 2017--2017 & 1 & subgroup & 1997--1997 & 1 \\
        angle-doppler & 2016--2016 & 1 & polyserial & 1996--1996 & 1 \\
        circuit-based & 2016--2016 & 1 & phi & 1995--1995 & 1 \\
        fuzzy interval & 2016--2016 & 1 & partial serial & 1980--1980 & 1 \\
        higher order & 2016--2016 & 1 & noise & 1970--1970 & 1 \\
        innovative & 2016--2016 & 1 & sample genetic & 1969--1969 & 1 \\
        Pearson's rank-variate & 2016--2016 & 1 & simple & 1969--1969 & 1 \\
        regression & 2016--2016 & 1 & linear & 1968--1968 & 1 \\
        wavelet & 2016--2016 & 1 & sine & 1968--1968 & 1 \\
        angles & 2015--2015 & 1 & Daniels' & 1961--1961 & 1 \\
        bivariate & 2015--2015 & 1 & non-circular serial & 1960--1960 & 1 \\
        cepstral & 2015--2015 & 1 & gross & 1956--1956 & 1 \\
        circular & 2015--2015 & 1 & polynomial & 1952--1952 & 1 \\
        frequency & 2015--2015 & 1 & interfunction & 1950--1950 & 1 \\
        head-weighted gap-sensitive & 2015--2015 & 1 & parabolic & 1946--1946 & 1 \\
    \end{tabular}
    \caption{\label{tab:corr_coef} Table of Correlation Coefficients from Literature Search (see details in text).}
\end{table}

\end{document}